# Breaking Kirchhoff's Law in Nonlinear Thermal Emission

Ruixin Ma[1], Yijia Yu[1], Yuncong Sun[1], Hengzhe Yan[1], Xianfeng Chen[2], Wenjie Wan[1,2*]

[1]State Key Laboratory of Photonics and Communications, University of Michigan-Shanghai Jiao Tong University Joint Institute,
Shanghai Jiao Tong University, Shanghai 200240, China
[2]Department of Physics and Astronomy, Shanghai Jiao Tong University, Shanghai 200240, China
*Corresponding author: wenjie.wan@sjtu.edu.cn

**Thermal radiation is strictly governed by Kirchhoff's law to reach thermal equilibrium. The violation of Kirchhoff's law decouples nonreciprocally the equity between absorptivity and emissivity, enabling exotic thermal engineering applications. However, achieving broadband nonreciprocal thermal emissivity and absorptivity remains a challenge. Here we experimentally demonstrate nonreciprocal and broadband thermal radiation by breaking Kirchhoff's law through nonlinear optical frequency conversion in a scattering medium. Thermal blackbody radiation is upconverted through sum-frequency generation with an intense infrared pump, while broadband conversion is enabled by the critical random quasi-phase-matching condition in the nonlinear nanocrystals. Moreover, a temporal transient measurement also indicates a possible active radiation cooling through such nonlinear thermal radiation. These results may pave a new way for nonlinear and active thermal management in critical applications like radiation cooling, energy harvesting, and infrared camouflage.**

Thermal emission, the natural process by which all objects with a finite temperature radiate electromagnetic waves, is fundamentally important to energy technologies. Traditionally, thermal emission adheres to the principle of reciprocity, where an object's absorbed and emitted radiation are equal for a given wavelength. This fundamental equality was formalized by Gustav Kirchhoff in 1860. Based on that, advancements in metamaterials [1-3], photonic/phononic crystals [4-7], surface plasmon/phonon polaritons [8-10] offer a powerful tool to tailor thermal emission with defined spectral [11,12], directional [13,14], and dynamic characteristics [15-17]. This law has profoundly influenced the design of thermal emitters for centuries. However, recent breakthroughs in nonreciprocal thermal radiation are challenging this long-standing equality of Kirchhoff's Law [18]. By breaking Lorentz reciprocity, it becomes possible to decouple absorptivity and emissivity, offering exciting new possibilities for applications such as active radiation cooling [19-22], energy harvesting [23-26], heat transfer [27-29], and infrared sensing/camouflage [30,31].

The reciprocity is hard to break, but nonreciprocal systems manifest themselves through magnetic fields [32,33], nonlinearity [34-36], and time modulation [37-39]. To achieve a violation of Kirchhoff's law, a pioneer work using structured magneto-optical materials has demonstrated thermal emitters can exhibit nonreciprocal directional emission/absorption under strong external magnetic fields [40-42]. In addition, topological materials such as Weyl semimetal films have been studied for the purpose of achieving nonreciprocal thermal emitters [43,44]. Space-time-modulated metasurfaces have also been theoretically proposed to show a strong contrast between absorptivity and emissivity without magnetic fields [45,46]. It is well-known that nonlinearity can also break time-reciprocity, especially in optics, such nonlinear and nonreciprocal devices are of critical importance for on-chip optical isolators [47,48], optical circulators [49]. The nonlinear approach to the thermal emission may offer a magnetic-field free, broadband, wide-angle, and polarization-independent solution to the nonreciprocal thermal emission [50-53]. Although nonlinear optical frequency conversions with thermal emission have been theoretically proposed in [54], and experimentally realized in [55,56], none of them has addressed the critical issue of violation of Kirchhoff's law till now.

In this work, we experimentally demonstrate a violation of thermal Kirchhoff's law using a GaAs nanocrystalline scattering medium of quadratic nonlinearity, where emissivity and absorptivity at the same wavelength diverge due to the presence of sum-frequency generation (SFG). An intense infrared pump at 10.6μm induces a broadband frequency conversion of thermal radiation, enabled by random quasi-phase matching in the nanocrystals. The resulting upconverted emission significantly alters the radiation spectrum around its peak, where sum-frequency generation enhances emissivity controlled by the pump. Meanwhile, the nonlinear optical gain of propagating mid-infrared light leads to a net reduction in apparent absorptivity, breaking the equity between the emissivity and absorptivity. Furthermore, we demonstrate tunable control over radiative heat dissipation by the pump in a transient measurement. These findings provide new nonlinear approaches to the fundamental limits of thermal reciprocity and open pathways toward actively controlled thermal management for various applications in radiation cooling, energy harvesting, and infrared sensing/camouflage.

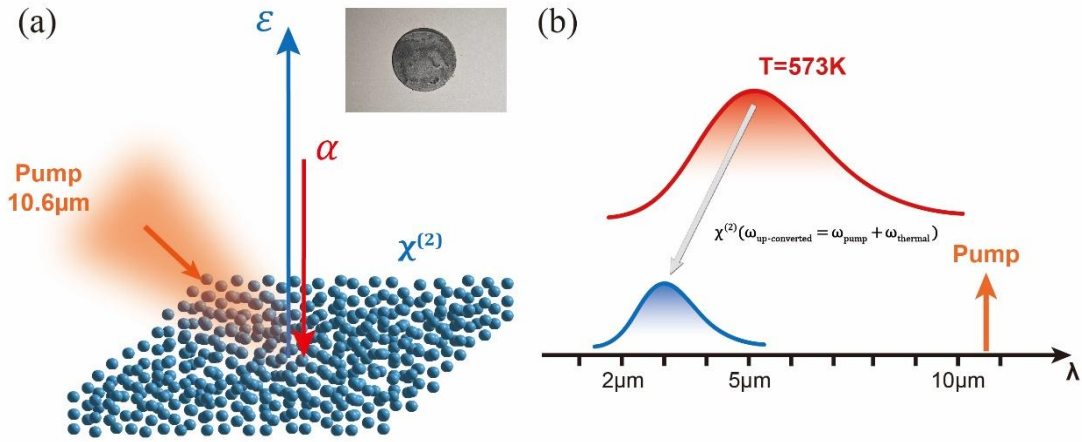

**Figure 1: The violation of thermal Kirchhoff's law through nonlinear frequency conversion.** (**a**) Under optical pumping, the GaAs nanocrystals, with strong second-order nonlinearity, exhibit nonreciprocal thermal radiation behavior: the emissivity increases while the absorptivity decreases. (**b**) Such effect originates from broadband nonlinear frequency conversion between different wavelengths within the GaAs nanocrystals: the nonlinear upconversion of thermal radiation directly enhances the emissivity, while the probe light inside the nanocrystals experiences a gain from nonlinear conversion.

In general, for non-magnetic, time-invariant, and linear systems, the emissivity and absorptivity of an object are equal at a given wavelength, termed Kirchhoff's law of thermal radiation. However, in a nonlinear system, nonlinearity-induced nonreciprocity can break this equivalence due to nonlinear frequency conversion between different wavelengths. Figure 1 illustrates the underlying principle of violating the thermal Kirchhoff's law through nonlinear frequency conversion. As shown in Fig. 1(b), when a thermal emitter with a strong second-order nonlinear susceptibility $\chi^{(2)}(1/\lambda_{up-conv}=1/\lambda_{pump}+1/\lambda_{ther})$ is at temperature T, its intrinsic thermal radiation is primarily concentrated around wavelength $\lambda_{ther}$. Upon the illumination of a strong external pump at wavelength $\lambda_{pump}$, a portion of the thermal radiation is upconverted to wavelength $\lambda_{up-conv}$ via random quasi-phase-matching (RQPM) [55,56]. This nonlinear frequency conversion directly enhances the emissivity of the material at $\lambda_{up-conv}$ through the following formula:

$$\Delta\varepsilon(\lambda) \propto \left\langle \left|\chi^{(2)}\right|^2 \right\rangle \cdot I_{ther}(\lambda_{ther},T)I_{pump} > 0 \qquad (1)$$

The change in emissivity, denoted as $\Delta\varepsilon$, arises from nonlinear thermal radiation within the nanocrystalline medium, which is proportional to the intensities of both the pump beam and the material's intrinsic thermal radiation. Here, $\left\langle \left|\chi^{(2)}\right|^2 \right\rangle$ is the averaged second-order nonlinear coefficients contributed from all geometric orientations under RQPM condition, as detailed in Supplementary Materials S1 [59].

In the meantime, when a weak probe beam passes through a nanocrystalline medium with a strong second-order nonlinear susceptibility, it experiences not only the intrinsic linear absorption of the material but also an additional nonlinear gain. This gain originates from the interplay between the material's second-order nonlinearity and the externally applied pump field. That is:

$$\Delta\alpha(\lambda) \propto -\left\langle \left|\chi^{(2)}\right|^2 \right\rangle \cdot \eta \cdot I_{ther}(\lambda_{ther}, T) I_{pump} < 0 \tag{2}$$

Here it is worth noting that the presence of the probe wave will affect the efficiency of sum-frequency generation inside the nanocrystalline medium, introducing an additional correction factor $0 < \eta < 1$ in Equation (2). The correction factor approaches unity under weak probe assumption, which leads to $\Delta\varepsilon(\lambda, T) = -\Delta\alpha(\lambda, T)$. When no external pump is applied, the system remains linear, and the emissivity and absorptivity are equal according to Kirchhoff's law. However, once the pump is turned on, the nonlinear nanocrystals exhibit enhanced emissivity, but reduced absorptivity at a specific wavelength band, indicating a breakdown of this equality.

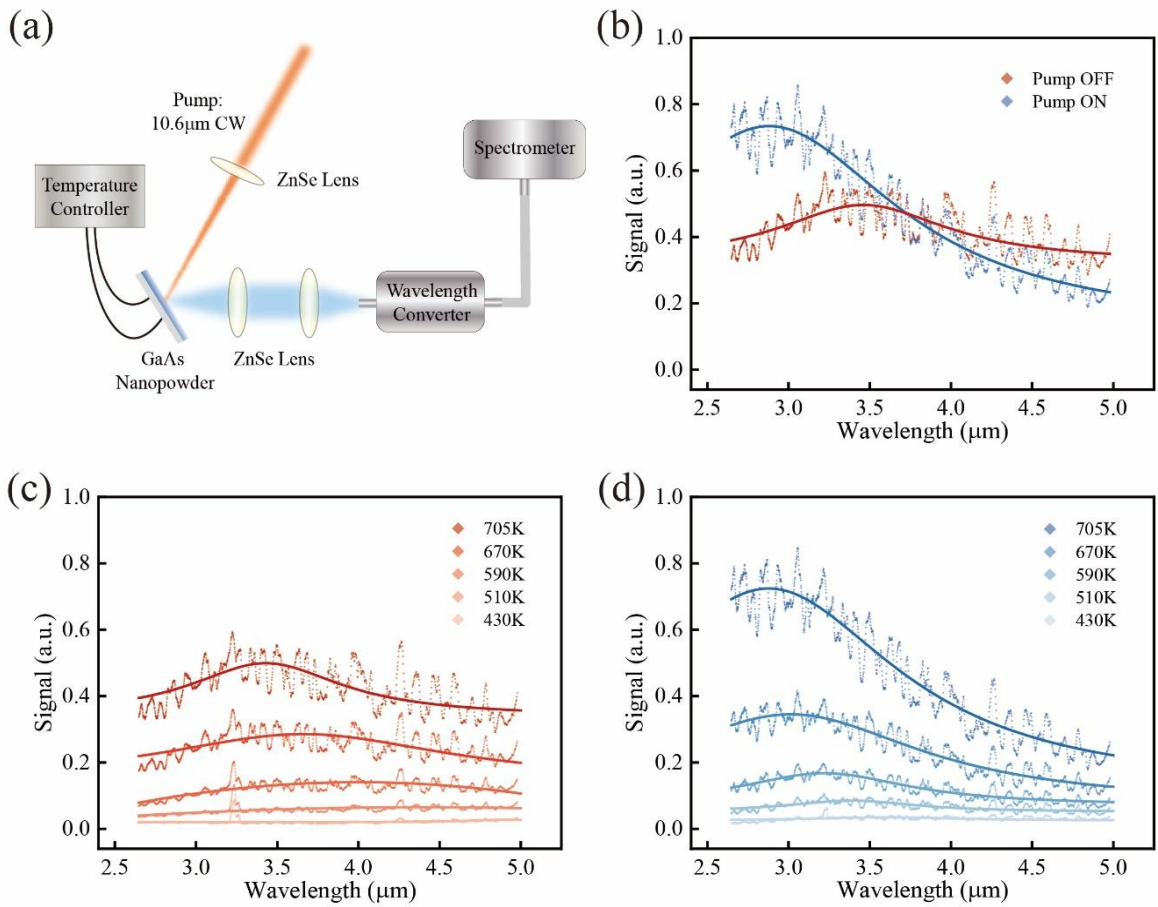

**Figure 2: A broadband nonlinear frequency conversion of thermal radiation.** (**a**) Schematic of the experimental setup. A CW pump at 10.6μm is focused onto the GaAs nanocrystals and the resulting mid-infrared thermal signal is collected and then sent through a wavelength converter, followed by detection with a high-resolution spectrometer. (**b**) Measured emission spectra at 705 K with the pump on and off, showing spectral modification and confirming efficient nonlinear conversion under optical pumping. (**c**)(**d**) Temperature-dependent spectra for the unpumped (red curves) and pumped (blue curves) cases, revealing enhanced short-wavelength radiation under pumping.

Experimentally, Gallium Arsenide (GaAs) nanocrystals are used as nonlinear thermal emitters due to their broad transparency window, spanning from 1μm to 16μm, and strong second-order nonlinearity [57]. Within this spectral range, GaAs exhibit relatively low and weakly wavelength-dependent absorption coefficients. The GaAs nanocrystal thin films are fabricated through grinding method [55]. When the GaAs nanocrystal samples are heated to approximately 300 °C, their thermal emission is predominantly centered near 5μm (Fig. 1). Through sum-frequency generation, the upconverted wavelength can be estimated as $\lambda_{up-conv} = (\lambda_{pump}^{-1} + \lambda_{ther}^{-1})^{-1} = 3.4\mu m$, corresponding to approximately 3.4μm. Consequently, a portion of the thermal radiation initially at 5μm is shifted to the 3.4μm region, resulting in enhanced emission in this spectral band. To ensure broadband frequency conversion, a random quasi-phase-matching (RQPM) scheme is implemented in such nonlinear and random scattering medium [54]. The optimal phase matching condition can be reached when the average GaAs particle size is approximately 84.8μm according to our theoretical calculation, see Supplementary Materials S7 [59].

Figure 2 provides spectral evidence of broadband nonlinear upconversion in the GaAs nanocrystals under infrared pumping. The experimental setup is shown in Fig. 2(a): a continuous-wave $CO_2$ laser at 10.6μm is focused onto the GaAs nanocrystalline medium using a ZnSe lens. Emitted mid-infrared radiation is collected by two ZnSe lenses and directed into a wavelength converter, which maps the 2.5–5μm band to the 746–877 nm range. The upconverted signal is detected with a high-resolution spectrometer, enabling analysis of the sample's thermal emission. To suppress substrate heating, the GaAs powder was deposited on a Ge substrate, which is transparent from 2 to 16μm [58]. The sample temperature was precisely controlled to vary thermal radiation levels. Fig. 2(b) shows spectra at 705 K with and without pumping, acquired under identical conditions (100s integration). A clear enhancement near 3μm is observed with the pump on, along with a slight reduction near 5μm — indicative of sum-frequency upconversion. Temperature-dependent spectra are also shown in Figs. 2(c) and 2(d) for the unpumped and pumped cases, respectively. Without pumping, the emission increases smoothly with temperature, lacking sharp features. Under pumping, a pronounced peak emerges near 3μm and grows with temperature, diverging from the unpumped spectral profile. These results confirm efficient nonlinear frequency conversion in the GaAs nanocrystalline scattering medium, redistributing thermal emission toward shorter wavelengths. This shift underlies the observed nonreciprocal thermal radiation in the system.

To show the violation of Kirchhoff's law, Figure 3 shows pump-induced modulation of emissivity and absorptivity in the GaAs nanocrystals. The setup is outlined in Fig. 3(a), with full details in the Supplemental Material S2 [59]. For absorptivity, a 3.3μm CW probe was used, and changes in transmitted intensity under varying pump powers revealed absorption modulation. Emissivity was measured by recording spontaneous emission near 3.3μm with the probe off. Fig. 3(b) shows the measured emissivity and absorptivity in the GaAs nanocrystals and Figs. 3(c) and (d) show the corresponding time-domain traces for emissivity and absorptivity, respectively. Clearly, the emissivity increases nearly linearly with pump power, while the absorptivity decreases with increasing pump power. This clear divergence between the emissivity and the absorptivity violates Kirchhoff's law. As expected from Kirchhoff's law, the emissivity and absorptivity converge when no pump is applied. The linear emissivity increase aligns with sum-frequency generation (SFG) scaling, consistent with prior observations in lithium niobate nanocrystalline scattering medium [55,56]. The reduction in absorptivity also trends linearly, suggesting the pump-induced SFG gain.

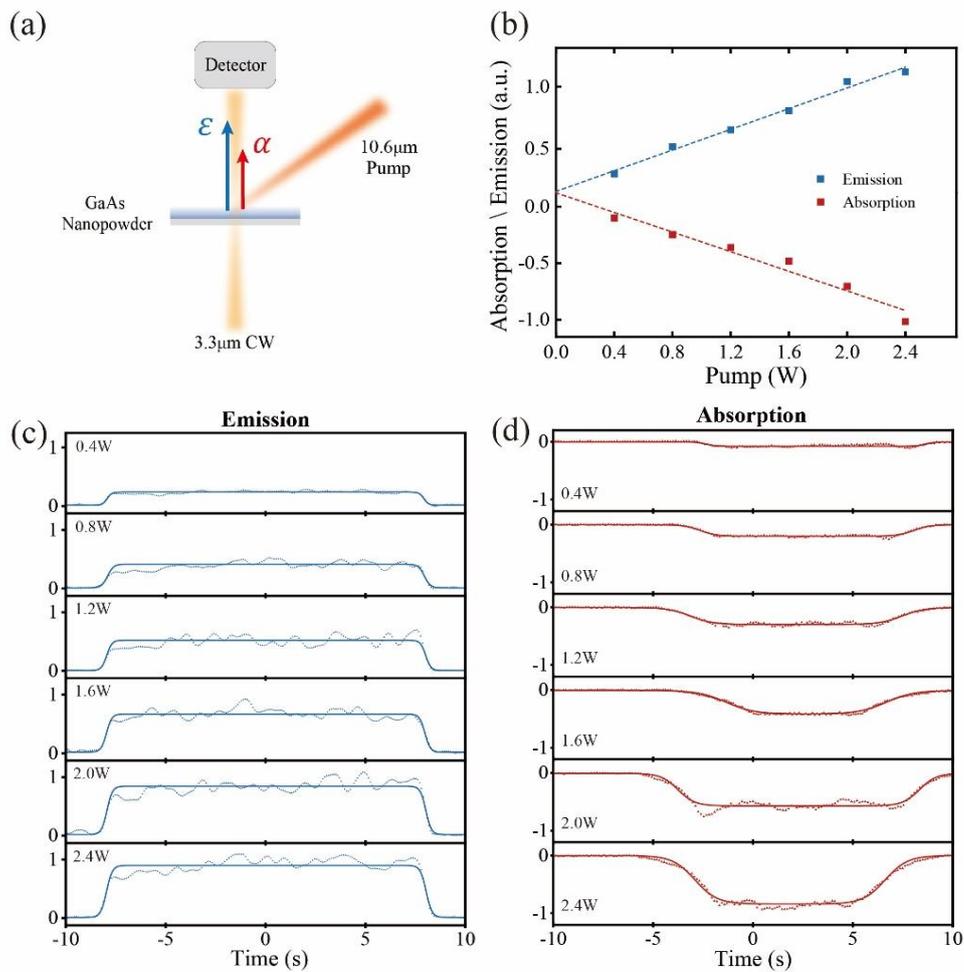

**Figure 3: Violation of Kirchhoff's law in pump-induced absorptivity and emissivity.** (**a**) Schematic diagram of the measurement setup. Emissivity is measured through spontaneous emission near 3.3μm with the probe off. And absorptivity is measured through the transmitted intensity of the weak probe beam. (**b**) Measured emissivity (blue) and absorptivity (red) in the GaAs nanocrystals under varying pump powers. The divergence shows clear evidence of violating the thermal Kirchhoff's law. (**c**)(**d**) Corresponding time-domain signals for emissivity (blue) and absorptivity (red), respectively. Transitions in the signals indicate the on/off switching of the pump.

At last, to demonstrate the possible radiative cooling through nonlinear thermal emission, Figure 4 illustrates how nonlinear thermal emission affects heat dissipation. As shown in Fig. 4(a), a temperature sensor, a thermistor composed of a platinum resistance wire, can typically dissipate heat through three conventional mechanisms: thermal conduction, thermal convection, and thermal radiation. Under optical pumping, the GaAs nanocrystal layer exhibits enhanced emissivity over absorptivity, introducing an additional nonlinear radiative dissipation channel. Fig. 4(b) presents the measured thermal equilibration time of temperature sensors as a function of pump power. Correspondingly, Fig. 4(c) and 4(d) show a comparison of temperature response curves for the bare and coated sensors, respectively. For the bare sensor, equilibration time remains nearly constant with increasing pump power, indicating that the heat dissipation rate is unaffected by the pump. In contrast, the GaAs-coated sensor shows a clear reduction in equilibration time, implying an enhanced thermal dissipation pathway. This behavior is attributed to nonlinear upconversion in the GaAs nanocrystals layer, which increases its effective emissivity under optical pumping. At low pump powers, the bare and coated sensors already exhibit different equilibration times, likely due to differences in surface morphology and thermal contact. Importantly, numerical simulations confirm that increases in pump alone do not significantly alter equilibration time, supporting the radiative origin of the observed effect, see Supplementary Materials S3 [59].

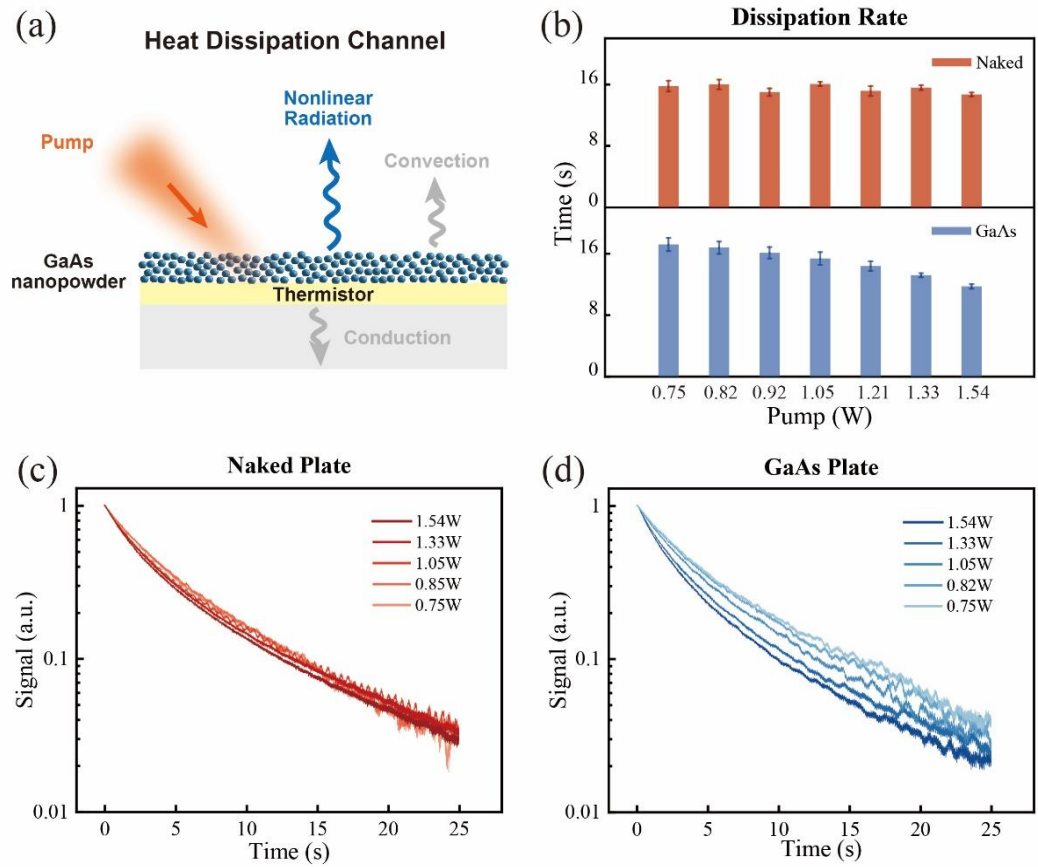

**Figure 4: Heat dissipation through nonlinear thermal radiation.** (**a**) The schematic diagram for heat dissipation pathways. Under optical pumping, an additional nonlinear radiative dissipation channel is introduced for a GaAs-coated sample. (**b**) Comparison of thermal equilibration times under varying pump powers for heat sources with and without GaAs nanocrystal coatings. (**c**)(**d**) Normalized temperature-related signals in the time domain for bare (red) and coated (blue) sensors, implying an enhanced thermal dissipation pathway.

In summary, we have experimentally demonstrated nonreciprocal thermal radiation in a GaAs nanocrystalline scattering medium and shown its ability to modulate heat dissipation. Unlike conventional magneto-optical approaches, this nonlinear optical method requires no magnetic materials and allows tunability of emission and absorption spectra via pump intensity and wavelength. The intrinsic random quasi-phase matching (RQPM) mechanism enables broadband operation, supporting multiband nonreciprocal thermal emission. While the nonlinear signals are weaker than those in magneto-optical systems and require sensitive lock-in detection, their broadband nature offers a distinct advantage. These findings open new avenues for manipulating radiative heat flow and may stimulate further research in nonlinear heat transfer and nonequilibrium thermodynamics.

**Acknowledgment**: This work was supported by the National Science Foundation of China (Grant No. 12274295); Shanghai Technology Innovation Project (Grant No. 24590711300); and the National Key Research and Development Program (No. 2023YFA1407200; Grant No. 2023YFB3906400).

**Data:** The data that support the findings of this article are openly available [60].